\documentstyle[preprint,aps]{revtex}
\begin{document}
\draft
\def\pmb#1{\setbox0=\hbox{#1}%
     \kern-.025em\copy0\kern-\wd0
      \kern.05em\copy0\kern-\wd0
       \kern-.025em\raise.0433em\box0}
\def\btau{\pmb{$\tau$}}
\def\bsigma{\pmb{$\sigma$}}
\def\bdiamond{\pmb{$\diamond$}}
\def\bbdiamond{\pmb\bdiamond}
\def\hat{\widehat}
\def\omeg{\bmatdieci\char '41}
\title
{Coulomb effects on growth of instabilities\\ in 
asymmetric nuclear matter}

\author{G. Fabbri and F. Matera} 

\address
{Dipartimento di Fisica, Universit\`a degli Studi di Firenze\\
and Istituto Nazionale di Fisica Nucleare, Sezione di
Firenze,\\
L.go E. Fermi 2, I-50125, Firenze, Italy}

\maketitle
\begin{abstract}
~We study the effects of the Coulomb interaction 
on the growth of unstable modes in asymmetric nuclear matter. 
In order to compare with previous calculations we use 
a semiclassical approach based on the linearized Vlasov 
equation. Moreover, a quantum calculation is performed 
within the R.P.A.. The Coulomb effects are a slowing down 
of the growth and the occurrence of a minimal 
wave vector for the onset of the instabilities. The quantum 
corrections cause a further decrease of the growth rates. 
\end{abstract}

\pacs{PACS number(s): 21.65.+f, 21.60.Ev, 21.60.Jz, 25.70.Pq}

Very recently there has been a growing number of studies on 
the properties of highly asymmetric nuclear matter. This can be 
mainly ascribed to recent experimental investigations on 
processes involving neutron--rich nuclei \cite{exp}.  

Under certain kinematic conditions, a nuclear collision 
can produce physical situations in which bulk or surface 
instabilities take place. 
Several theoretical models have been proposed in order to
study such situations \cite{gro90,mor93,bon95}. Among microscopic 
approaches, that
based on the Boltzmann--Langevin equation is one of the most
suitable to describe the large fluctuations observed in
fragment formation \cite{+al,cata}. In this scheme the instabilities
of the self--consistent mean field can play a crucial role
in enhancing mass and charge fluctuations \cite{coldito}.

In reactions between neutron-rich nuclei regions of high charge 
asymmetry, like the neck zone in semi--peripheral
collisions, can be produced. In these regions
the  chemical (~or diffusive~) instability can be the most effective
\cite{bao97}.

The thermodynamics of phase separation in asymmetric nuclear 
matter was first treated in studies concerning the structure 
of neutron stars \cite{bay71}. 
More recently thermodynamic approaches have been developed 
with specific attention to critical situations that can be 
reached in collisions of heavy ions \cite{bao97,mu95}. 

From thermodynamics one can obtain both the equilibrium 
configurations and the conditions for the onset of the chemical 
instability, i. e. the critical values of neutron and
proton densities and of temperature for a given equation of state.
Thus it is possible to draw the borders of the instability 
in the pressure--density plane for instance. 
The result is that for asymmetric nuclear matter the 
region of  mechanical instability is embedded in the region of 
chemical instability \cite{bao97}. This means that in asymmetric 
nuclear matter the effective spinodal region is defined by the 
chemical instability. 

In this report, rather than with the thermodynamics of phase 
separation, we are concerned with the dynamical development of the 
instabilities in asymmetric nuclear matter. The growth rates of the 
various unstable modes in asymmetric nuclear matter have been already 
studied in Refs.~\cite{fab97,bar97} using a Skyrme force for the 
effective nucleon--nucleon interaction. Here we want to evaluate also 
the effects of the Coulomb force on the growth of instabilities. 
It is well known that the Coulomb force gives a divergence in the 
energy density of infinite nuclear matter. Thus it 
cannot be taken into account in thermodynamic studies of such a 
physical system. However we will see that within a mean--field treatment 
of nuclear excitations, the contribution from the Coulomb 
force does not give rise to divergences. Moreover, we assume
that the unperturbed initial state is uniform and homogeneous.   
In this case the mean field of the initial 
state does not appear in calculations explicitly. Only the 
temperature, the density and the asymmetry of the initial state 
actually occur in the relevant equations. 

In the study of the nuclear dynamics by means of kinetic equations, 
the self--consistent mean field is usually treated in semiclassical 
approximation \cite{+al,cata,coldito}. In order to compare 
with previous calculations, here we study  the unstable modes of 
asymmetric nuclear matter by using the linearized Vlasov equation. 
This equation is a semiclassical approximation of the RPA, valid in 
the long--wavelenght limit. However, we also evaluate the quantum 
corrections to the semiclassical results. 

Since the RPA is a linear approximation, we expect that the validity 
of these calculations is limited to times close enough to the 
onset of instabilities. An estimate of the time interval, in which the 
RPA can be considered valid, is given in Ref.~\cite{bar97}. 
There, the numerical solutions of the non linear Vlasov equation 
have been compared to the analytical results of 
the linear approximation. The two procedures give quite similar 
results for times $t\lesssim 150 fm/c$. 

In our calculations we have used a Skyrme--like form for the nuclear 
part of the nucleon--nucleon effective interaction. We start from 
the following simplified  functional for the density of the 
nuclear potential energy: 
\begin{equation}
{\cal E}^{(N)}(\varrho_1,\varrho_2)=\,\frac{A}{2}\frac{\varrho^2}{\varrho
_{eq}}+\frac{B}{\sigma+2}\,\frac{\varrho^{\sigma+2}}{\varrho_{eq}^{\sigma+1}}
+\frac{D}{2}(\nabla\varrho)^2+\frac{C}{2}\frac{\varrho_A^2}{\varrho_{eq}}
-\frac{D^\prime}{2}(\nabla\varrho_A)^2~,
\label{nuclen}
\end{equation}
where $\varrho=\varrho_1+\varrho_2$ is the total density,
$\varrho_1$ and $\varrho_2$ are the proton and neutron densities 
respectively, $\varrho_{eq}$ is the density of symmetric nuclear 
matter at saturation  and $\varrho_A=\varrho_2-\varrho_1$. 
For the parameters in Eq.~(\ref{nuclen}) we take the 
values:                                                          
\[
A=-356.8\,MeV,~~B=303.9\,MeV,~~\sigma=\,\frac{1}{6}~,
\]
\[ 
C=32\,MeV,~~D=130\,MeV\cdot fm^5,~~D^\prime =40\,MeV\cdot fm^5~.
\]
The values of $A$, $B$ and $\sigma$ reproduce the binding energy 
(~$15.75\,MeV$~) and give an incompressibility modulus of $201\,MeV$ 
for symmetric nuclear matter at saturation with 
$\varrho_{eq}=0.16\,fm^{-3}$. For the values of $D$ and $D^\prime$ 
we follow the prescriptions of Ref.~\cite{mye66} and 
Ref.~\cite{bay71} respectively. Finally this value of the parameter 
$C$ gives a symmetry energy coefficient of $28\,MeV$ in the 
Bethe--Weizs\"acker mass formula.  The energy density of 
Eq.~(\ref{nuclen}) 
coincides with that used in Ref.~\cite{bar97}, except for the symmetry 
energy term for which we prefer the simplest required form.

Concerning the Coulomb energy density, we use the expression given 
by the Hartree--Fock approximation, with the Fock term 
evaluated in the local density approximation: 
\begin{equation}
{\cal E}^{(C)}({\bf r})=\frac{e^2}{2}\varrho_1({\bf r})\,
\int d{\bf r}^\prime\frac{\varrho_1({\bf r}^\prime)}
{\vert {\bf r}-{\bf r}^\prime\vert}-\frac{3}{4}\Bigl(\frac{3}{\pi}\Bigr)
^{\frac{1}{3}}e^2\varrho_1^{\frac{4}{3}}~.\label{coulen}
\end{equation}

Within the density functional theory the potential energy functional 
\begin{equation}
E^{(pot)}[\varrho_1,\varrho_2]=\int d{\bf r}\,\bigl({\cal E}^{(N)}
+{\cal E}^{(C)}
\bigr)
\end{equation}
is the quantity, which allows to define an effective nucleon--nucleon 
interaction for R.P.A. calculations \cite{bra93}. 
The interaction is given by  the double functional 
derivative of $E^{(pot)}[\varrho_1,\varrho_2]$ 
\begin{equation}
V_{i,j}({\bf r},{\bf r}^\prime)=\,\frac{\delta^2 E^{(pot)}}
{\delta\varrho_i({\bf r})\delta\varrho_j({\bf r}^\prime)}\bigg|_0
\label{pot}
\end{equation}
calculated in the unperturbed initial state. As 
expected, the Hartree term of the Coulomb energy yields the bare 
Coulomb force, whereas the remaining terms of Eq.~(\ref{nuclen}) and the 
exchange Coulomb term give rise to zero--range forces. 

The R.P.A. equations for the oscillations of the proton and 
neutron densities, $\delta\varrho_1({\bf r},t)$ and 
$\delta\varrho_2({\bf r},t)$, around their unperturbed values, 
can be obtained by standard procedures (~e.g. see Ref.~\cite{pines66}~). 
Here we report the resulting equations for their space and time 
Fourier transforms. For a wave of frequency $\omega$ 
and wave--vector ${\bf k}$, the following two 
coupled equations are obtained
\begin{mathletters}
\label{allrpa}
\begin{eqnarray}
&\bigg(&1-\Phi^{(T)}_1(k,\omega)\Big(F_V+k^2(D+D^\prime)
+\frac{C}{\varrho_{eq}}+\frac{4\pi e^2}{k^2}
+V_C^{(ex)}\Big)\bigg)\delta\varrho_1(k,\omega)
\nonumber \\
&-&\Phi^{(T)}_1(k,\omega)\bigg(F_V+k^2(D-D^\prime)-
\frac{C}{\varrho_{eq}}\bigg)\delta\varrho_2(k,\omega)=\,0~,
\label{rpa.a}
\end{eqnarray}
\begin{eqnarray}
&-&\Phi^{(T)}_2(k,\omega)\bigg(F_V+k^2(D-D^\prime)-\frac{C}{\varrho_{eq}}
\bigg)\delta\varrho_1(k,\omega)
\nonumber \\
&+&\bigg(1-\Phi^{(T)}_2(k,\omega)\Big(F_V+k^2(D+D^\prime)
+\frac{C}{\varrho_{eq}}\Big)\bigg)\delta\varrho_2(k,\omega)=\,0~,
\label{rpa.b}
\end{eqnarray}
\end{mathletters}
where 
\begin{equation}
F_V=\,\frac{A}{\varrho_{eq}}+(\sigma+1)B\,\frac{\varrho_0^{\sigma}}
{\varrho^{\sigma+1}_{eq}}
\end{equation}
is the volume term of the effective nuclear interaction and 
\begin{equation}
V_C^{(ex)}=\,-\frac{1}{3}\Big({\frac{3}{\pi}}\Big)^{\frac{1}{3}}e^2
\varrho_{01}^{-\frac{2}{3}}
\end{equation}
is the exchange term of the Coulomb interaction. The quantities 
$\varrho_0$ and $\varrho_{01}$ denote the total and proton 
densities in the unperturbed initial state. In Eqs.~(\ref{allrpa}) 
$\Phi^{(T)}_i(k,\omega)$ represents the free particle--hole 
propagator, which in the long--wavelenght limit takes the form 
\begin{equation}
\Phi^{(T)}_i(k,\omega)=\,-\int d\epsilon_p\,\Phi(p,k,\omega)\, 
\frac{\partial n_i}{\partial\epsilon_p}
~,\label{propa}
\end{equation}
where
\begin{equation}
n_i=\,\frac{2}{e^{(\epsilon_p-\tilde\mu_i)/T}+1}
\end{equation}
is the mean occupation number of protons or neutrons in the state 
with kinetic energy $\epsilon_p=p^2/2m$
and
\begin{equation}
\Phi(p,k,\omega)=\,\frac{1}{(2\pi)^3}\int d\Omega_{\bf p}
\frac{{\bf v}\cdot{\bf k}}{\omega+i\eta-{\bf v}\cdot{\bf k}}~,
\end{equation}
${\bf v}=\,{\bf p}/m$ is the nucleon velocity. We use units such 
that $\hbar=1,~c=1$. 
The effective chemical potential $\tilde\mu_i$ in the distribution 
$n_i$, is the chemical potential measured with 
respect to the uniform mean field acting on the nucleons 
in the unperturbed initial state. For a given temperature $T$ it is 
completely determined by the nucleon densities. Therefore 
the ingredients of Eqs.~(\ref{allrpa}) related to the unperturbed initial 
state are only the neutron and proton densities, besides the 
temperature. Moreover, we remark that the product 
$\Phi^{(T)}_1(k,\omega)4\pi e^2/k^2$ does not diverge when $k\to 0$. 

The dispersion relation $\omega=\,\omega(k)$ is obtained by 
equating to zero the determinant of the set of equations (\ref{allrpa}). 
We look for solutions of Eqs.~(\ref{allrpa}) for values of $\varrho_0,~T$
and the asymmetry parameter $\alpha=(\varrho_{02}-\varrho_{01})
/\varrho_0$ inside
the region of instability \cite{bao97}.
In general we find that 
the frequency $\omega(k)$ is or purely imaginary or purely real. 
The growth rate of the instability is given by $\Gamma(k)=\,Im\,\omega(k)$. 

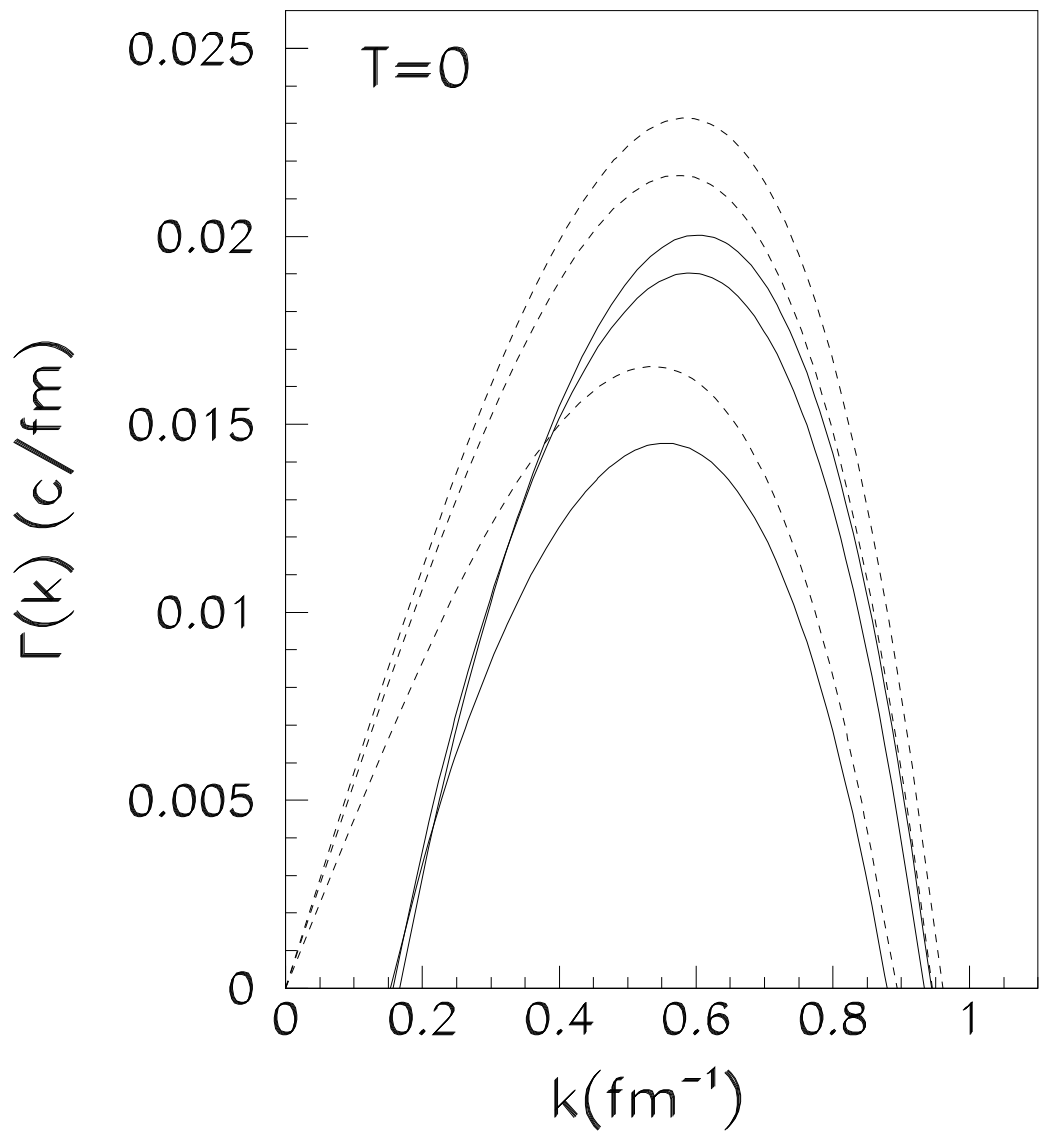
\begin{figure}
\caption{Growth rates of the unstable modes calculated in the 
semiclassical approximation \Big(~Eqs.~(\ref{allrpa})~\Big) for 
$T=0$ and for three different values of the asymmetry parameter 
$\alpha$. From top to bottom $\alpha=0.0,0.3,0.6$. The 
density of nuclear matter is $\varrho_0=\,0.4\varrho_{eq}$ . 
Results with the complete interaction (~solid lines~) and  
without the Coulomb interaction (~dashed lines~). 
} 

\label{f:1}
\end{figure}

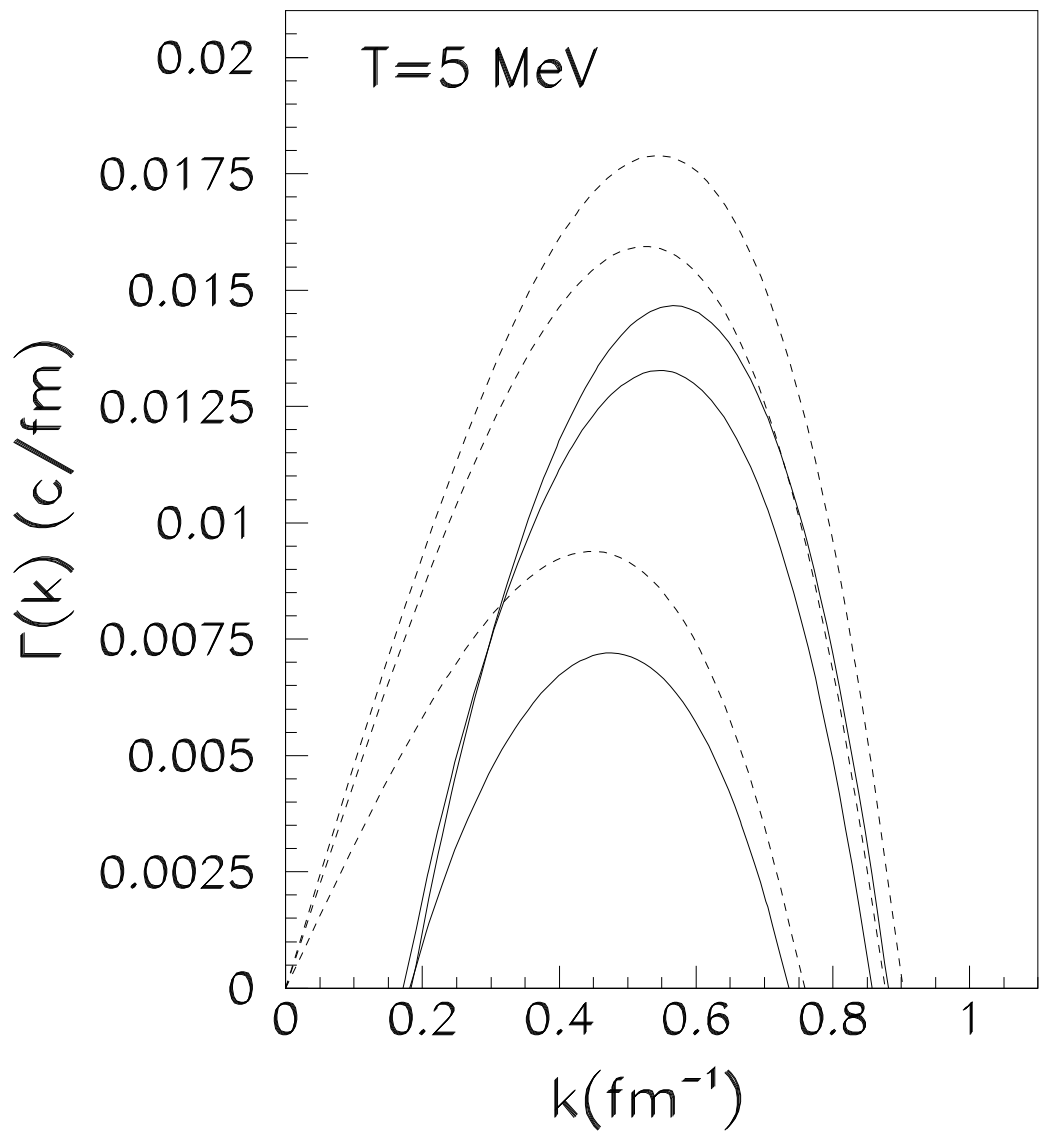
\begin{figure}
\caption{The same as Fig.~\ref{f:1}, but for $T=5\,MeV$
} 
\label{f:2}
\end{figure}

In Figs.~\ref{f:1} and \ref{f:2} the calculated growth rate is 
displayed as a function of $k$ for $\varrho_0=\,0.4\,\varrho_{eq}$ 
and for various values of the parameters $\alpha$ and $T$. 
Figures~\ref{f:1} and \ref{f:2} show that the Coulomb force causes an 
overall decrease of the growth rate $\Gamma(k)$. This decrease  
depends only sligthly on the temperature(~$\sim 10\%$~) and it is 
almost independent of the asymmetry. Moreover we observe that, when the Coulomb 
force is included, the wave vector $k$ must exceed 
a certain value $k_{min}$ in order to get solutions of Eqs.~(\ref{allrpa}) 
with $\Gamma(k)\not=\,0$. Below  $k_{min}$ the solutions 
of Eqs.~(\ref{allrpa}) correspond to undamped plasmon--like oscillations, 
which practically involve the proton density alone. In the present case 
$k_{min}$ is about $1/5$ of the Fermi momentum of the symmetric 
nuclear matter (~$\alpha=0$~). These two effects (~decrease of 
$\Gamma(k)$ and appearance of $k_{min}$~) are due to the 
competition between the Coulomb and nuclear forces 
(~responsible for the chemical instability~). The Coulomb 
force pushes the protons towards regions of lower density, the 
nuclear forces instead push the neutrons in the same direction.  
Explicit calculations, with and without the Coulomb force, 
confirm that the proton and neutron densities oscillate according 
to this picture. 

So far we have presented  results of calculations performed in the 
semiclassical limit of R.P.A.. For infinite systems this limit 
essentially coincides with the Vlasov equation. However, 
Figs.~\ref{f:1} and \ref{f:2}  show that the values of the wave vector 
of the most unstable modes is not a small fraction of the proton 
and neutron Fermi momenta (~for $\alpha=\,0,~k/p_F\sim 0.6$~). 
Therefore, an evaluation of the quantum corrections to the 
Vlasov equation is of interest. Moreover, since for increasing 
values of $\alpha$ the Fermi momentum of the protons decreases 
appreciably, the importance of the quantum corrections can 
depend on the asymmetry degree of the nuclear matter. For 
infinite systems and for the nucleon--nucleon interaction  
used here, it is a rather simple matter to perform full quantum 
RPA calculation. It is sufficient to substitute 
in Eqs.~(\ref{allrpa}) the semiclassical version of the free 
particle--hole propagator given by Eq.~(\ref{propa}) with its 
full quantum expression. This can be obtained by a straightforward 
generalization to finite temperature of the expression for $T=\,0$ 
given in Ref.~\cite{fetter71}. 

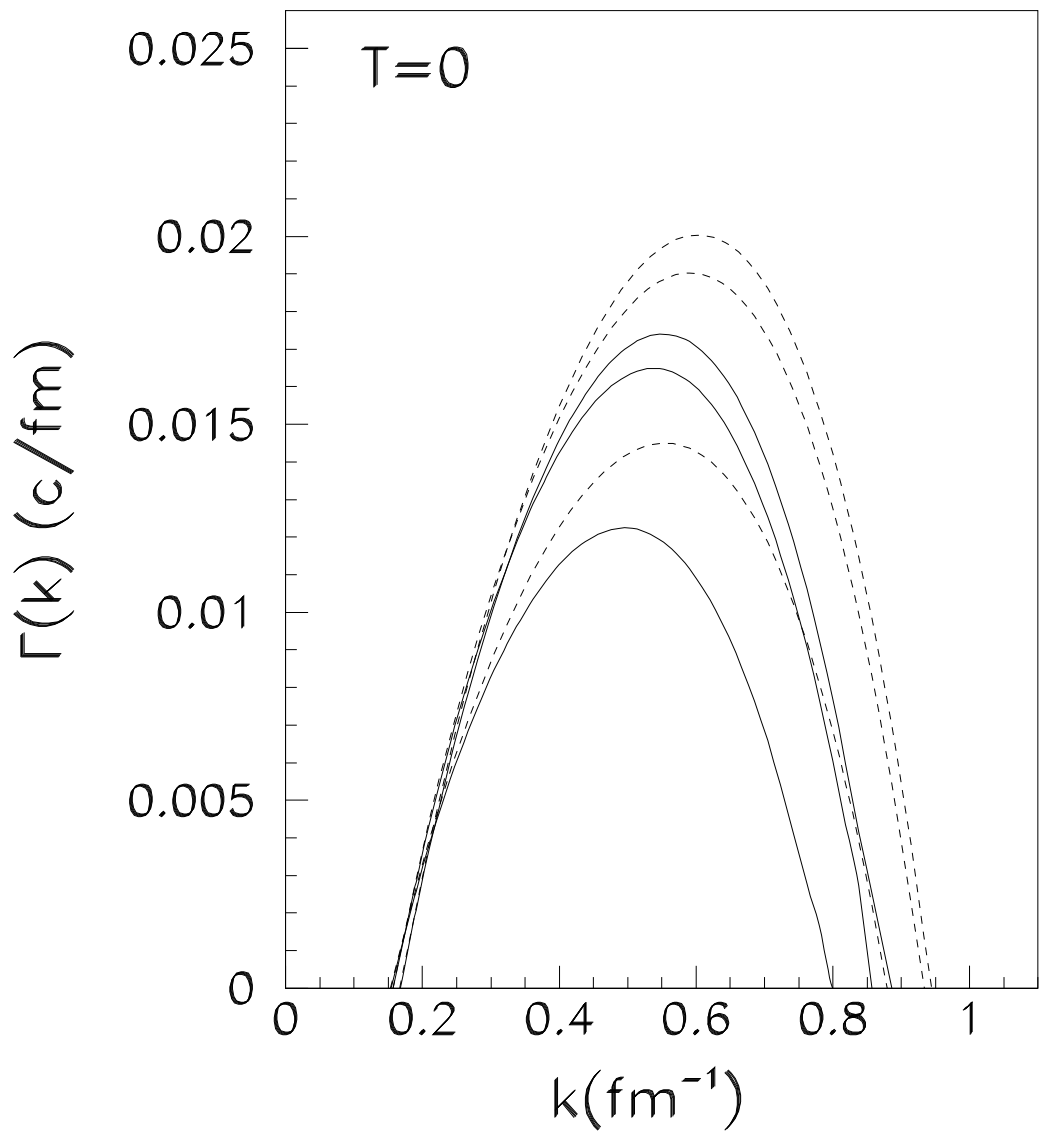
\begin{figure}
\caption{Growth rates of the unstable modes, calculated with 
the quantum corrections (~solid lines~) and in the semiclassical 
approximation (~dashed lines~). The values of the parameters are the 
same as in Fig.~\ref{f:1}. 
} 

\label{f:3}
\end{figure}

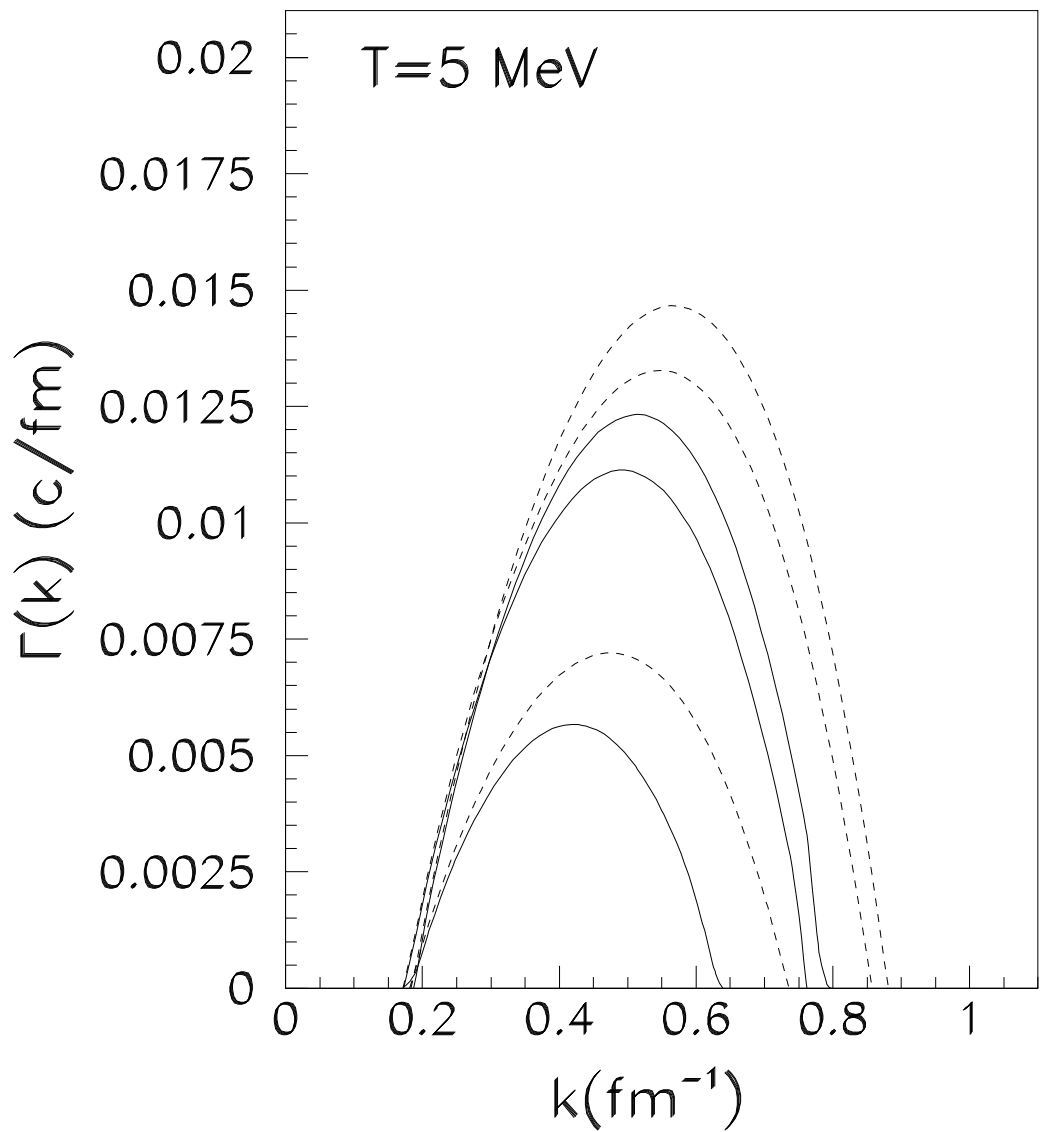
\begin{figure}
\caption{The same as Fig.~\ref{f:3}, but for $T=5\,MeV$. 
} 
\label{f:4}
\end{figure}

In Figs.~\ref{f:3} and \ref{f:4}  we show the effects of the 
quantum corrections on the growth rates of the unstable modes. 
We observe a further quenching of the dispersion curves 
$\Gamma(k)$. To be more specific, the maxima of $\Gamma(k)$ 
are lowered by a factor of $\sim 0.85$ for all 
the values of temperature and asymmetry parameter $\alpha$ shown.

For symmetric nuclear matter, a similar evaluation of quantum 
effects has been done in Ref.~\cite{ay95}, without the Coulomb 
interaction and with a different nuclear interaction.  
Neglecting the Coulomb interaction, for $\alpha=\,0$ we obtain   
results in qualitative agreement with those of Ref.~\cite{ay95}. 

In conclusion, for the growth rate of the unstable modes  the net 
result of the Coulomb effects and of the quantum corrections 
can be summarized by a suppression factor of $\sim 0.7$, which does 
not significantly change in the range of temperature and asymmetry 
considered here. Moreover, the long range character of the Coulomb 
force inhibits the formation of chemical instabilities 
of wavelength $\lambda >2\pi/k_{min}$ (~$\simeq 32\,fm$ for
$\varrho_0=\,0.4\varrho_{eq}$~). 

\acknowledgements

We are grateful to Dr. A. Dellafiore for valuable discussions 
and for a careful reading of the manuscript.

\end{document}